\def\paperversion{normal}

\def\paperversiondraft{draft}
\def\paperversionnormal{normal}

\ifx\paperversion\paperversionnormal
  \documentclass[sigplan, screen, nonacm]{acmart}
\else
  \def\paperversion{draft}
  \documentclass[sigplan, screen, review, nonacm]{acmart}
\fi

\makeatletter
\if@ACM@anonymous
  
\else
  
\fi
\makeatother

\usepackage{colortbl}

\usepackage{environ}
\ifx\paperversion\paperversiondraft
  
\else
  \NewEnviron{draftonly}{}
\fi

\usepackage[utf8]{inputenc}
\usepackage[T1]{fontenc}
\usepackage{microtype}

\usepackage{alltt}
\usepackage{xargs}
\usepackage{lipsum}
\usepackage{xparse}
\usepackage{xifthen, xstring}
\usepackage{xspace}
\usepackage{marginnote}
\usepackage{etoolbox}
\usepackage[acronym,shortcuts]{glossaries}
\usepackage{glossaries-extra}
\usepackage{amsmath}
\usepackage{algorithm}
\usepackage[noend]{algpseudocode}
\usepackage{hyphenat}
\usepackage[shortcuts]{extdash}
\usepackage{subcaption}
\usepackage[inline]{enumitem}
\usepackage{multirow}
\usepackage{tikz}
\usepackage[frozencache]{minted}

\AtEndPreamble{\usepackage{thmtools}} 

\usetikzlibrary{shadows.blur, quantikz2}

\definecolor{pairedNegOneLightGray}{HTML}{cacaca}
\definecolor{pairedNegTwoDarkGray}{HTML}{827b7b}
\definecolor{pairedOneLightBlue}{HTML}{a6cee3}
\definecolor{pairedTwoDarkBlue}{HTML}{1f78b4}
\definecolor{pairedThreeLightGreen}{HTML}{b2df8a}
\definecolor{pairedFourDarkGreen}{HTML}{33a02c}
\definecolor{pairedFiveLightRed}{HTML}{fb9a99}
\definecolor{pairedSixDarkRed}{HTML}{e31a1c}
\definecolor{dark-orange}{HTML}{d95f02}

\colorlet{light-red}{pairedFiveLightRed!40}
\colorlet{light-green}{pairedThreeLightGreen!40}
\newminted[xdsl]{tools/lexers/MLIRLexer.py:MLIRLexer}{mathescape, autogobble, escapeinside=££, style=murphy}
\newminted[xdsl-top]{tools/lexers/MLIRLexer.py:MLIRLexer}{mathescape, autogobble, escapeinside=££, style=murphy, highlightcolor={light-red}, highlightlines={3,4,6,8}}
\newminted[xdsl-bottom]{tools/lexers/MLIRLexer.py:MLIRLexer}{mathescape, autogobble, escapeinside=££, style=murphy, highlightcolor=light-green, highlightlines={6,7,8}}
\newmintinline[xdslinline]{tools/lexers/MLIRLexer.py:MLIRLexer}{mathescape, escapeinside=££, style=murphy}

\setabbreviationstyle[acronym]{long-short}
\glssetcategoryattribute{acronym}{nohyperfirst}{true}

\newacronym{ir}{IR}{Intermediate Representation}
\newacronym{ssa}{SSA}{Static Single-Assignment}
\newacronym{mbqc}{MBQC}{Measurement-Based Quantum Computing}
\newacronym{qec}{QEC}{Quantum Error Correction}

\graphicspath{{./images/}}

\makeatletter
\newcommand\requiredelimiter[2][########]{%
  \ifdefined#2%
    \def\@temp{\def#2#1}%
    \expandafter\@temp\expandafter{#2}%
  \else
    \@latex@error{\noexpand#2undefined}\@ehc
  \fi
}
\@onlypreamble\requiredelimiter
\makeatother

\newcommand\newdelimitedcommand[2]{
  \expandafter\newcommand\csname #1\endcsname{#2}
  \expandafter\requiredelimiter
  \csname #1 \endcsname
}

\newdelimitedcommand{toolname}{Tool}

\usepackage{booktabs}

\usepackage[verbose]{newunicodechar}
\newunicodechar{₁}{\ensuremath{_1}}
\newunicodechar{₂}{\ensuremath{_2}}
\newunicodechar{∀}{\ensuremath{\forall}}
\newunicodechar{α}{\ensuremath{\alpha}}
\newunicodechar{β}{\ensuremath{\beta}}

\newcommand{\writegate}[1]{\relax\ifmmode\mathsf{#1}\else\textsf{#1}\fi\xspace}
\newcommand{\X}{\writegate{X}}
\newcommand{\Y}{\writegate{Y}}
\newcommand{\Z}{\writegate{Z}}
\renewcommand{\H}{\writegate{H}}
\renewcommand{\S}{\writegate{S}}
\newcommand{\T}{\writegate{T}}
\newcommand{\CX}{\writegate{CX}}
\newcommand{\CZ}{\writegate{CZ}}
\newcommand{\Id}{\writegate{I}}
\newcommand{\XZ}{\writegate{XZ}}
\newcommand{\XZS}{\writegate{XZS}}
\newcommand{\XY}{\writegate{XY}}
\newcommand{\R}{\writegate{R}}
\newcommand{\J}{\writegate{J}}
\newcommand{\U}{\writegate{U}}

\begin{document}

\title[A Dynamic Intermediate Representation for Hybrid Quantum-Classical Programs]{A Dynamic Intermediate Representation\\for Hybrid Quantum-Classical Programs}       


\author{Alex Rice}
\orcid{0000-0002-2698-5122}
\affiliation{
  \institution{University of Edinburgh}
  \country{United Kingdom}
}
\email{alex.rice@ed.ac.uk}

\author{Chris Heunen}
\orcid{0000-0001-7393-2640}             
\affiliation{%
  \institution{University of Edinburgh}           
  \country{United Kingdom}
}
\email{chris.heunen@ed.ac.uk}         

\author{Tobias Grosser}
\orcid{0000-0003-3874-6003}             
\affiliation{
  \institution{University of Cambridge}            
  \country{United Kingdom}
}
\email{tobias.grosser@cst.cam.ac.uk}          

\settopmatter{printfolios=true}

\begin{abstract}
  Quantum compilers typically follow the circuit model, representing programs as fixed sequences of gates. This static view breaks down in hybrid quantum-classical applications, where gate choices depend on runtime data or measurement results. We introduce a new Intermediate Representation (IR) that elevates gates to first-class values, enabling their dynamic creation, composition, and control. This unified representation allows classical computation to steer quantum behaviour, capturing phenomena including stochastic gate selection, adaptive error correction, and measurement-driven computation within a single framework. Case studies in noise modelling, randomised compilation, error correction, and measurement-based quantum computing show that our IR expresses these programs compactly and supports optimisations that were not possible in the circuit model. Evaluation on a benchmark suite of hybrid quantum-classical programs indicates that our IR represents programs compactly and facilitates compiler analysis and transformation.
\end{abstract}

\maketitle

\section{Introduction}

Quantum circuits are the dominant representation of quantum computations~\cite{divincenzoMesoscopicElectronTransport1997}.
However, the circuit model is inherently static and too restrictive for many important applications, which are more naturally expressed as hybrid quantum-classical programs combining quantum operations with classical data and control flow~\cite{raussendorfOneWayQuantumComputer2001,roffeQuantumErrorCorrection2019,corcolesExploitingDynamicQuantum2021,piveteausutter:knitting}. Established quantum toolkits have begun extending support for classical computation within quantum compilers, such as IBM’s OpenQASM 3.0~\cite{crossOpenQASM3Broader2022} and Quantinuum's t\(|\)ket\(\rangle\) 2~\cite{Sivarajah_TKET_A_Retargetable_2020}, emphasising the importance of representing subcircuits that depend on external classical or probabilistic computations.

Consider noisy quantum channels, which are essential for modelling real hardware~\cite{preskill:nisq}. The phase-flip channel, a common form of noise, probabilistically applies a \Z gate to a qubit. As the choice of gate depends on a classical random variable, the circuit model can only represent this behaviour by generating a fresh circuit for each random sample.

A more general representation of quantum computation should  support optimisation across the quantum-classical boundary, be expressive enough to encode diverse hybrid techniques, and remain simple enough for efficient compilation and transformation. Unlike classical hardware---like in an ASIC or FPGA, where circuit structure is fixed at runtime~\cite{takanoReconfigurationCostReconfigurable2022,rouxSurveyReducingReconfiguration2014}---quantum programs are executed dynamically, allowing circuit layout to change during execution. Thus there is no fundamental reason to adhere to a static circuit model, and instead we can adopt a dynamic representation better suited for optimising hybrid programs.

\begin{figure}[t]
  \centering
  \newsavebox\circuitone
  \sbox\circuitone{
    \begin{quantikz}
      &\gate{\Z}\wire[l][1]["q_0"{below}]{a}&\wire[l][1]["q_1"{below}]{a}
    \end{quantikz}
  }
  \newsavebox\circuittwo
  \sbox\circuittwo{
    \begin{quantikz}
      \ghost{\Z}&\phantomgate{\Z}\wire[l][1]["q_0"{below}]{a}&\wire[l][1]["q_1"{below}]{a}
    \end{quantikz}
  }
  \newsavebox\circuitthree
  \sbox\circuitthree{
    \begin{quantikz}
      &\gate{\Z}\wire[l][1]["q_1"{below}]{a}&\wire[l][1]["q_2"{below}]{a}
    \end{quantikz}
  }
  \newsavebox\circuitfour
  \sbox\circuitfour{
    \begin{quantikz}
      \ghost{\Z}&\phantomgate{\Z}\wire[l][1]["q_1"{below}]{a}&\wire[l][1]["q_2"{below}]{a}
    \end{quantikz}
  }
  \newsavebox\lhs
  \sbox\lhs{
    \begin{subfigure}[b]{\linewidth}
      \centering
      \begin{tikzpicture}[auto,
        box/.style = {draw,blur shadow,shadow blur radius=0.7ex, fill=white,align=center,on grid}]
        \node[box] (b1)    {Input \(q_0\)};
        \node[box, inner sep=0pt, below left = 1.3 and 2 of b1] (b2) {\usebox\circuitone};
        \node[box, inner sep=0pt, below right = 1.3 and 2 of b1] (b3) {\usebox\circuittwo};
        \node[box, inner sep=0pt, below = 1.7 of b2] (b4) {\usebox\circuitthree};
        \node[box, inner sep=0pt, below = 1.7 of b3] (b5) {\usebox\circuitfour};
        \node[box, below right = 0.8 and 2 of b4] (b6)    {Output \(q_2\)};
        \path[->] (b1) edge[bend right=20] node[pos=0.6,anchor=center,fill=white,inner sep=0.5pt]{\(p_0 = 1\)} (b2) ;
        \path[->] (b1) edge[bend left=20] node[pos=0.6,anchor=center,fill=white,inner sep=0.5pt]{\(p_0 = 0\)} (b3) ;
        \path[->] (b2) edge node[pos=0.5,anchor=center,fill=white,inner sep=0.5pt]{\(p_1 = 1\)} (b4) ;
        \path[->] (b3) edge node[pos=0.5,anchor=center,fill=white,inner sep=0.5pt]{\(p_1 = 0\)} (b5) ;
        \path[->] (b2.south east) edge[bend right=15] node[pos=0.25,anchor=center,fill=white,inner sep=0.5pt]{\(p_1 = 0\)} (b5.west) ;
        \path[->] (b3.south west) edge[bend left=15] node[pos=0.25,anchor=center,fill=white,inner sep=0.5pt]{\(p_1 = 1\)} (b4.east) ;
        \path[->] (b4) edge[bend right=20] (b6) ;
        \path[->] (b5) edge[bend left=20] (b6) ;
      \end{tikzpicture}
      \caption{With control flow}
      \label{fig-cf-vs-dyn-gate-control-flow}
    \end{subfigure}
  }
  \usebox\lhs
  \vfill
  \begin{subcaptionblock}[b]{0.49\linewidth}
    \centering
    \begin{quantikz}[column sep = 1em]
      &\setwiretype{n}p_0 \mathrel{?} \Z : \Id\arrow[d, shorten >= 2pt]&p_1 \mathrel{?} \Z : \Id\arrow[d, shorten >= 2pt]&\\
      &\gate[style={fill=none}]{\phantom{\Z}}\wire[l][1]["q_0"{below}]{a}&\gate[style={fill=none}]{\phantom{\Z}}\wire[l][1]["q_1"{below}]{a}&\wire[l][1]["q_2"{below}]{a}
    \end{quantikz}
    \caption{With \xdslinline{arith.select}}
    \label{fig-cf-vs-dyn-gate-dynamic-gates}
  \end{subcaptionblock}
  \begin{subcaptionblock}[b]{0.49\linewidth}
    \centering
    \begin{quantikz}
      &\gate{\X^0\Z^{p_0}}\wire[l][1]["q_0"{below}]{a}&\gate{\X^0\Z^{p_1}}\wire[l][1]["q_1"{below}]{a}&\wire[l][1]["q_2"{below}]{a}
    \end{quantikz}
    \caption{With \XZ gadgets}
    \label{fig-cf-vs-dyn-gate-xz-gadgets}
  \end{subcaptionblock}
  \caption{
    Graphical representations of two sequential phase flip channels using control flow in (a), and dynamic gates in (b) and (c), with the gate value being generated by \xdslinline{arith.select} and our \XZ gadget respectively.
    The dataflow is vastly simplified in (b) and (c), where the quantum component retains its circuit structure.
    The \XZ gadgets in (c) remove the need for the selection operation, allowing the dynamic gates to be fused.
  }
  \label{fig-cf-vs-dyn-gate}
\end{figure}

In this work, we introduce \emph{dynamic gates}, a representation for hybrid quantum-classical computation that treats gates as \emph{runtime values}. Rather than defining a separate operation for each quantum gate, a gate value can be created, manipulated by classical code, and applied dynamically during execution.

\autoref{fig-phase-flip} illustrates this concept using two consecutive phase-flip noise channels. The key construct, \xdslinline{qssa.dyn_gate}, takes as input a dynamically chosen gate value \xdslinline|

\begin{listing}[t]
\begin{xdsl}
\end{xdsl}
  \caption{
    Our dynamic gates can represent two concatenated probabilistic phase flip channels without requiring control flow.
    Quantum-classical transformations can then reduce this to a single phase flip with updated probability, using the self-invertibility of the quantum \Z gate.
  }
  \label{fig-phase-flip}
\end{listing}

Dynamic gates make the boundary between quantum and classical computation porous, enabling optimisations that cross this boundary while preserving the familiar circuit structure. They can be viewed as a natural generalisation of the circuit model, and can replace instances of more explicit control flow necessary in other IRs~\cite{QIRSpec2021,crossOpenQASM3Broader2022}.

  \noindent Our contributions are:
  \begin{itemize}[topsep=2pt, leftmargin=15pt]
  \item A \emph{dynamic gate model} for tightly coupled hybrid quantum-classical programs, instantiated as a compiler \ac{ir} that enables optimisations across the quantum-classical interface (\autoref{sec-dyn-gates}, \autoref{sec-ir}).
  \item The \emph{\XZ gadget}, a dynamic gate representing conditional combination of \X and \Z Pauli gates allowing an elegant way to realise optimisations (\autoref{sec-xz}).
  \item A curated \emph{benchmarking suite} of hybrid programs that is representative of current demands on quantum-classical interaction.
  \end{itemize}
We show that this \ac{ir} advances the state of the art in representing and optimising hybrid quantum-classical programs in two ways (\autoref{sec-case-studies}):
\begin{itemize}[topsep=2pt, leftmargin=15pt]
\item Depth: Through case studies in \emph{randomised compilation} (\autoref{sec-rand-comp}), \emph{quantum error correction} (\autoref{sec-qec}), and \emph{measurement-based quantum computing} (\autoref{sec-mbqc}), we demonstrate how complex hybrid transformations become simple local rewrites.
\item Breadth: Through evaluation on our new benchmark suite, covering nontrivial quantum-classical interactions and scalable program sizes, we demonstrate that our \ac{ir} compactly expresses hybrid programs and enables comparative analysis across representations (\autoref{sec-benchmarking}).
\end{itemize}

\section{Dynamic Gates}
\label{sec-dyn-gates}

Designing \acp{ir} for classical computation is a well-studied problem, as is designing \acp{ir} to represent the quantum circuit model.
A naive union of a classical \ac{ir} and quantum circuit \ac{ir} can therefore represent both quantum and classical computation, but struggles to model the interaction between the two components, which is integral for many modern quantum computing applications~\cite{raussendorfOneWayQuantumComputer2001,roffeQuantumErrorCorrection2019,corcolesExploitingDynamicQuantum2021,piveteausutter:knitting}.

An example of an optimisation that crosses the boundary between the classical and quantum components of the program is fusing consecutive quantum phase flip channels (\autoref{fig-cf-vs-dyn-gate}).
In our dynamic gate model this can be represented as a single circuit, whereas it becomes a complicated branching of circuits when mixing classical and quantum \acp{ir}.

A phase flip channel acts on a single qubit, and is equivalent to conditionally applying a \Z gate to the qubit with some probability \(p\).
Phase flip channels may not immediately appear to be hybrid quantum-classical programs, but representing them requires quantum operations to depend on the result of classical computation, in this case the instantiation of a random variable. Thus they demonstrate the tools we have developed well.

\begin{figure*}
  \begin{minipage}[][][t]{0.42\linewidth}
\begin{xdsl-top}

\end{xdsl-top}
    \subcaption{Double phase flip with \XZ gadgets.}
    \label{fig-xz-double-phase}
  \end{minipage}
  \begin{minipage}[][][t]{0.42\linewidth}
\begin{xdsl-bottom}

\end{xdsl-bottom}
    \subcaption{Double phase flip after fusion.}
    \label{fig-xz-fused}
  \end{minipage}
  \caption{
    The \XZ gadget allows the phase flip channel to be represented without any selection operations~(a).
    This allows the consecutive dynamic gates to be fused, reducing the program to a single phase flip channel~(b).
  }
  \label{lst-xz-phase}
\end{figure*}

Performing two such channels in sequence is equivalent to performing a single one (with an updated probability):
with probability $(1-p)^2$ no \Z gates are applied,
with probability $p^2$ two \Z gates are applied which cancel out,
and with probability $2p(1-p)$ a single \Z gate is applied.
Such a transformation should be simple to perform in a compiler, but note that the necessary transformation is not purely quantum or purely classical.
Thus, how easy it is to perform such a simplification depends on the quantum-classical interface exposed by the \ac{ir}.

One way to represent the channel is to use classical control flow, either as blocks of computation linked by branching operations, or a more structured \texttt{if} conditional statement (\autoref{fig-cf-vs-dyn-gate-control-flow}).
Such an approach is not ideal for the transformation described above, as performing it would require reasoning across multiple control flow blocks.
Furthermore, this representation obscures the quantum dataflow of the program, because the same qubit can be used in multiple execution branches.
This complicates simple analyses. For example, to check if a qubit is used linearly, one must check that no execution trace uses the qubit twice. Verifying this property is undecidable in general and common over-approximations require global control flow analysis.

Instead, the dynamic gate model bridges the classical and quantum components of a program.
The fundamental primitive is the \emph{dynamic gate} operation.
Unlike the usual representation of gates in a quantum circuit, the gate to be run is supplied by an operand, instead of a static piece of information.
This input can be the result of arbitrary computation, providing the interface for classical data to affect quantum computation.
We refer to operands that specify quantum gates as \emph{gate values}.

\autoref{fig-phase-flip} represents two sequential phase flips in the dynamic gate model using our \ac{ir}. The gate value is obtained from a selection operation on a randomly generated boolean.
This shows the strength of the dynamic gate model: no complexity is added to the quantum dataflow by its interaction with classical dataflow.
The output qubit from the first phase flip passes directly to the input of the second phase flip.
Essentially, the quantum component of the program retains the structure of a quantum circuit (\autoref{fig-cf-vs-dyn-gate-dynamic-gates}).

The dataflow is simpler, but two classical choices of quantum gates still need to be fused.
In this small-scale example a full case analysis is possible, but would quickly become intractable in general.
To remedy this we introduce the \emph{\XZ gadget}, a succinct way to represent conditional combinations of Pauli \X and \Z gates.
So far, all gate values have been constant or classical selections between other constant gate values, but the dynamic gate model can be extended by further operations for generating gate values.
Therefore the \XZ gadget is implemented as a fresh operation that takes two boolean values \(x\) and \(z\) as operands and produces a gate value representing the gate \(\X^x\Z^z\).

The full utility of \XZ gadgets derives from the transformations which apply to it.
A selection operation between two \XZ gadgets can be replaced by a single \XZ gadget with selection operations on each of its operands.
Following this transformation with simple arithmetic simplifications reduces the double phase flip to the representation in \autoref{fig-xz-double-phase}, as illustrated in \autoref{fig-cf-vs-dyn-gate-xz-gadgets}.

Moreover, \XZ gadgets can be easily fused together by taking an exclusive disjunction of their parameters, which preserves the semantics of the gate value up to global phase.
This reduces our example to a single phase flip channel, given in \autoref{fig-xz-fused}.
As the \XZ gadget considers \X gates in addition to \Z gates, the same reduction sequence can similarly simplify compositions of bit-flip and depolarising channels.
\autoref{sec-case-studies} demonstrates that this gadget applies much more widely than the scenario presented here, because conditional Pauli operators are pervasive in quantum computing.

\section{Background}
\label{sec-background}

Now that the key idea of the paper has been introduced, let us step back to recall relevant background. We briefly review compilation, especially in the context of the MLIR framework, and the relevant parts of quantum computation.

\subsection*{Quantum computing}

The basic data in quantum computing is carried by \emph{qubits}. The state of a one-qubit system is represented by a vector in $\mathbb{C}^2$. Qubits subsume bits, with 0 and 1 being represented by the unit vectors 
$\ket 0 := \left(\begin{smallmatrix}1\\0\end{smallmatrix}\right)$ and 
$\ket 1 := \left(\begin{smallmatrix}0\\1\end{smallmatrix}\right)$.

Computations on qubits are represented by \emph{unitary} linear maps $\mathbb{C}^{2^n} \to \mathbb{C}^{2^n}$, that is, invertible matrices whose inverse is equal to their conjugate transpose. Especially, the \emph{Pauli gates} \X, \Y, and \Z satisfy some useful relations, such as
\[
  \X^2=\Y^2=\Z^2=-i\X\Y\Z=\Id=\left(\begin{smallmatrix}1&0\\0&1\end{smallmatrix}\right)
\]
and $\Z\X=i\Y=-\X\Z$.

Multiple qubits can be combined to form a register of qubits using \emph{tensor product}: the state of an $n$-qubit system is represented as a vector in $\mathbb{C}^2 \otimes \cdots \otimes \mathbb{C}^2 = (\mathbb{C}^2)^{\otimes n} \simeq \mathbb{C}^{2^n}$.
Important two-qubit gates are the controlled \X and \Z gates, denoted \CX and \CZ. These two gates, along with the Pauli gates, the Hadamard gate \H, and the phase gate \S are known as \emph{Clifford} gates, the gates that map Pauli gates to Pauli gates under conjugation. For example, \S satisfies the following equations:
\[ \Z\S = \S\Z \qquad \X\S = -\S\Y \qquad \Y\S = \S\X \]

Gates can be combined into \emph{quantum circuits}. They can be combined in parallel (with tensor products) and in sequence (with matrix multiplication). Quantum circuits form the \emph{de facto} standard hardware-agnostic low-level language for quantum computing. They are usually drawn graphically. For example, the quantum circuit
\begin{center}
  \begin{quantikz}[row sep = small]
    & \gate{\H} & \ctrl{1} & \meter{}  &\setwiretype{c} \\
    &          & \targ{}  & \gate{\S} &
  \end{quantikz}
\end{center}
represents a 2-qubit operation that performs an \H gate on the first qubit, a \CX gate on the first and second qubits, and then an \S gate on the second qubit.

The last operation on the first qubit is a \emph{measurement}, used to convert qubits into bits. This is a probabilistic operation. When measuring a qubit in state $\alpha v_1 + \beta v_2$ with respect to a basis \(\{v_1,v_2\}\), the resulting bit will be 0 with probability $|\alpha|^2$, and 1 with probability $|\beta|^2$.
Two states \(u\) and \(v\) are said to differ by a \emph{global phase} if \(v = e^{i\theta}u\) for some angle \(\theta\). Such states cannot be discriminated by measurement, and are hence identified.

To aid optimisation, quantum compilers often allow quantum circuits to include \emph{quantum gadgets} in addition to gates. These gadgets typically represent commonly occurring subcircuits or patterns within circuits, but are easier to simplify than the equivalent sequences of primitive gates. A representative example are \emph{phase gadgets}~\cite{cowtanPhaseGadgetSynthesis2020}; optimisation proceeds by identifying subcircuits that correspond to these phase gadgets, fusing adjacent gadgets together, and finally lowering each resulting gadget to primitive gates.

For many applications of quantum computing (see for example our case studies and benchmark suite in \autoref{sec-case-studies}), we must move beyond statically defined quantum circuits to \emph{hybrid} quantum programs. These hybrid programs include both classical and quantum computation, and can decide which quantum operations to perform at runtime, possibly depending on the results of measurements. It is not always trivial to translate optimisations from the simpler circuit setting to hybrid programs, and the hybrid setting admits new optimisations, especially those that concern both the quantum and classical components of the program.

For more information on quantum computing, we refer the reader to~\cite{nielsenQuantumComputationQuantum2010,yanofskymannucci:quantumcomputing}.

\subsection*{Compilers}

A compiler converts a computation from one representation to another, typically from a higher-level program to some form of machine code.
The term compiler is often used to refer to an optimising compiler, which attempts to minimise various metrics of the target code, such as program size or execution time.
Instead of performing a direct translation, compilers often transform a program into different \acp{ir}, each of which eases performing certain optimisations or analyses.

It is common for \acp{ir} to contain programs in Static Single-Assignment (SSA) form~\cite{rosenwegmanzadeck:ssa}, where each identifier is assigned a value in exactly one location. This makes dataflow explicit, as it is trivial to determine the origin of any value and to obtain a list of its uses.

To put a quantum gate in SSA form, it should consume the qubits it acts on, producing new fresh qubits as output. We contrast this \emph{value semantics} view of qubits to the \emph{reference semantics} view of qubits commonly used in existing quantum \acp{ir}, such as QASM and QIR, where a quantum gate has no outputs and is understood to be mutating its input qubits. The value of SSA form for optimisation is immediate; consider an optimisation that cancels gates with their inverses. To apply this to a single-qubit gate \(\U\) with input \(q\), it suffices to examine the unique origin of \(q\). Under reference semantics, \(q\) may have been mutated by any number of previously applied gates, complicating this analysis.

The \ac{ir} introduced in this paper is specifically presented as a set of MLIR~\cite{lattnerMLIRScalingCompiler2021} dialects, implemented on top of xDSL~\cite{fehrXDSLSidekickCompilation2025}, a Python clone of MLIR sharing the same textual representation.
Dialects in MLIR allow the user to add new types and operations to craft their own intermediate representations, while still having access to core data structures and optimisation machinery.

All \ac{ir} snippets in this paper follow MLIR syntax in style. Consider the following operation.
\begin{xdsl}
\end{xdsl}
The names \texttt{arith} and \texttt{qu} are names of dialects, which contain \texttt{select} and \texttt{bit} respectively, with the exclamation mark denoting that \xdslinline|!qu.bit| is an MLIR type. Any symbol beginning with \% is an SSA value, with operands to the left of the equality symbol and outputs to the right. This operation takes a boolean input \xdslinline|

Extra data can be attached to MLIR operations via attributes. Any symbol preceded by a \# is an attribute, for example \xdslinline|#gate.id| in \autoref{fig-phase-flip}.

\section{A Hybrid Quantum-Classical IR}
\label{sec-ir}

The \ac{ir} represents quantum operations in SSA form, with qubits having value semantics.
It is realised as a set of MLIR dialects. For completeness, we include the types, attributes, and operations used in the paper and the benchmark suite (see \autoref{sec-benchmarking}) in \autoref{tab:dynamic-gate-ir}.

In contrast to previous SSA-based quantum \acp{ir}~\cite{ittahQIROStaticSingle2022,peduriQSSASSAbasedIR2022,mccaskeynguyen:qir}, we do not introduce a new operation for each quantum gate, but instead add a single operation \xdslinline{qssa.gate}, for executing (static) gates.
The specific gate to be run is supplied by a \emph{gate attribute}, a static piece of compile-time data.
Gate attributes specify the number of qubits they act on, enabling verification of the operands and outputs to \xdslinline{qssa.gate}.

Specifying gates via attributes has multiple advantages.
Users can easily extend the gates available by introducing custom gates through new dialects.
Gates added in this way automatically cooperate with \xdslinline{qssa.gate} and its transformations.
Furthermore, the set of available gates can be reused between different applications.
Our implementation uses this to define a \texttt{qref} dialect in parallel to the \texttt{qssa} dialect, which represents quantum circuit operations with reference semantics instead of value semantics.
Available gates can be used in both dialects, and transformations are given between the two dialects without the need to inspect the gate attribute.

Most importantly, the various gate attributes can be reused to build our dynamic gate infrastructure.
 The dynamic gate operation \xdslinline{qssa.dyn_gate} (and its corresponding \texttt{qref} operation) takes the gate to be executed as an operand instead of an attribute.
This operand must have type \xdslinline{!gate.type<£\(n\)£>} for some \(n\), the type of \(n\)-qubit \emph{gate values}, runtime values storing an \(n\)-qubit quantum gate.

Gate values are primarily generated with \xdslinline{gate.constant}, which creates a gate value from a gate attribute.
Passing a constant gate value directly into a dynamic gate operation is equivalent to applying the corresponding static gate operation, and we give this simplification as a canonicalisation pattern on each respective dynamic gate operation.

As shown in \autoref{fig-phase-flip}, combining constant gate values with classical selection operations already allows more expressivity than the circuit model.
Applying a classically selected gate value is equivalent to branching with an \xdslinline{scf.if} operation.
The compiler pass \texttt{lower-dyn-gate-to-scf} performs this transformation.
Further lowering to basic block control flow gives a representation that is compatible with existing representations -- such as QIR~\cite{QIRSpec2021}, which inherits LLVM's control flow.

\begin{table}
  \centering
  \caption{
    Common gates as \XZ or \XZS gadgets.
    The \texttt{convert-to-xzs} compiler pass converts constant gate values of gates in the right hand column to \mbox{\XZ{}(\S)} gadgets.
    We assume \texttt{\%false} holds the constant boolean false, and \texttt{\%true} holds the constant true.
  }
  \label{tab-xz-gate}
  \begin{tabular}{cl}
    \toprule
    \textbf{Gate} & \textbf{Gadget Representation}\\
    \midrule
    \Id & \xdslinline|gate.xz  
    \X & \xdslinline|gate.xz  
    \Y & \xdslinline|gate.xz  
    \Z & \xdslinline|gate.xz  
    \S & \xdslinline|gate.xzs 
    \S{}\makebox[0pt][l]{\({}^\dagger\)} & \xdslinline|gate.xzs 
  \bottomrule
  \end{tabular}
\end{table}

Parallel with dynamic gates, the \ac{ir} contains a similar system for quantum measurements: a static measurement operation, \xdslinline{qssa.measure}, takes a measurement attribute; and a dynamic measurement operation, \xdslinline{qssa.dyn_measure}, takes a \emph{measurement value} as an operand.
In contrast to dynamic gates, a measurement attribute can be omitted, defaulting to a measurement in the computational basis.

The IR also adds tools for working with angles (represented as multiples of \(\pi\) quotiented by \(2\pi = 0\)), and operations for generating (classical) randomness.

\subsection{Gadgets for \X, \Z, and Phase (\S) Gates}
\label{sec-xz}

Non-constant gate values can also be generated through specialised operations, extending the dynamic gate model; for example, operations such as \xdslinline{gate.dyn_rx} dynamically generate rotation gates parameterised by a runtime angle.

We further include two more operations for generating dynamic gates, the \XZS gadget \xdslinline{gate.xzs}, and the \XZ gadget \xdslinline{gate.xz}. These operations (and their transformations) play a pivotal role in our case studies (see \autoref{sec-case-studies}), and so we pause to discuss them in greater depth.
The \XZS gadget creates compositions of Pauli \X, Pauli \Z, and Phase gates, and its simplification, the \XZ gadget, introduced in \autoref{sec-dyn-gates}, omits the Phase component.

The \XZS gadget takes three boolean values \(x\), \(z\), and \(s\) as input, outputting a gate value representing the composition \(\X^x\Z^z\S^s\). The \XZ gadget has only two inputs for the inclusion of the Pauli \X and \Z gates.
Various gates can be translated to \XZS (or \XZ) gadgets by the pass \texttt{convert-to-xzs}. These translations are given by
\autoref{tab-xz-gate}. As we only consider gates up to global phase, we can soundly define the Pauli \Y gate as the composition of Pauli \X and Pauli \Z gates, meaning the \XZ gadget can represent all Pauli gates.
With \xdslinline|
{\RecustomVerbatimEnvironment{Verbatim}{BVerbatim}{}
  \begin{center}
\begin{xdsl}
gate.xzs 
\end{xdsl}
  \end{center}}
\noindent Canonicalisation converts the left expression to the right.

The inputs to \XZ and \XZS gadgets can be arbitrary SSA values, and not just constant values, allowing the gadget to represent conditionally applied gates, and enabling powerful optimisations.
These optimisations organise into two pipelines, \texttt{xzs-simplify} and \texttt{xz-propagation} (\autoref{fig-pipelines}).
Both pipelines convert (possibly conditional) Pauli and Phase gates to \XZS gadgets, perform some transformation on these generated gadgets, and then lower back to selections between constant gates.
The \texttt{xzs-simplify} pipeline fuses adjacent \XZS gadgets, whereas the \texttt{xz-propagation} pipeline pushes \XZ gadgets towards the end of the circuit while fusing them, performing Pauli propagation with conditional gates.

\begin{figure}
  \centering
  \begin{tikzpicture}[
    box/.style = {draw,blur shadow,shadow blur radius=0.7ex, fill=white,align=center,on grid, rounded corners},
    leftarr/.style = {color=pairedTwoDarkBlue, thick},
    rightarr/.style = {color=dark-orange, thick}
    ]
    \node[box](s1) {\texttt{convert-to-xzs}};
    \node[right=3 of s1] {};
    \node[box, below=1 of s1] (s2) {\texttt{xzs-select}};
    \begin{scope}[transform canvas={xshift=-5pt}]
      \path[->, leftarr] (s1) edge (s2);
    \end{scope}
    \begin{scope}[transform canvas={xshift=5pt}]
      \path[->, rightarr] (s1) edge (s2);
    \end{scope}
    \node[box, below left=1 and 1.5 of s2] (s3) {\texttt{xzs-fusion}};
    \path[->, leftarr, shorten >= 2pt] (s2) edge (s3);
    \node[box, below right=1 and 1.5 of s2] (s4) {\texttt{xz-commute}};
    \path[->, rightarr, shorten >= 2pt] (s2) edge (s4);
    \path (s3) to node[midway,sloped] {\Large\(\subset\)} (s4);
    \node[box, below right=1 and 1.5 of s3] (s5) {\texttt{canonicalise}};
    \path[->, leftarr, shorten >= 2pt] (s3) edge (s5);
    \path[->, rightarr, shorten >= 2pt] (s4) edge (s5);
    \node[box, below=1 of s5] (s6) {\texttt{lower-xzs-to-select}};
    \begin{scope}[transform canvas={xshift=-5pt}]
      \path[->, leftarr] (s5) edge (s6);
    \end{scope}
    \begin{scope}[transform canvas={xshift=5pt}]
      \path[->, rightarr] (s5) edge (s6);
    \end{scope}
    \matrix [below left] at (current bounding box.north east) {
      \node [fill=pairedTwoDarkBlue,label=right:\small\texttt{xzs-simplify}] {}; \\
      \node [fill=dark-orange,label=right:\small\texttt{xz-propagation}] {}; \\
    };
  \end{tikzpicture}
  \caption{Optimisation pipelines for \XZS gadgets. The left path, \texttt{xzs-simplify}, fuses all adjacent conditional Pauli and Phase gates.
The right path, \texttt{xz-propagation}, attempts to move Pauli gates towards the end of the program while performing fusion.}
  \label{fig-pipelines}
\end{figure}

\paragraph{Interaction with selection: \texttt{xzs-select}}

If the inputs to an \xdslinline{arith.select} operation are \XZS or \XZ gadgets, then this selection can be converted to a single \mbox{\XZ{}(\S)} gadget with parameters given by selection operations between the parameters of the original gadgets.
More concretely,
\begin{xdsl}
\end{xdsl}
is replaced by
\begin{xdsl}
\end{xdsl}
\noindent with similar conversions being given for \XZ gadgets.

While this appears to make the program more complex, it is often the case that selections between boolean values can be trivially simplified, often to constant values, reducing the total number of operations.
When preceded by the pass \texttt{convert-to-xzs} and combined with canonicalisation for \xdslinline{arith.select}, this pass converts any conditional Pauli or Phase gate to its canonical \mbox{\XZ{}(\S)} gadget representation. For an example, see \autoref{fig-xz-double-phase}.

\paragraph{Merging sequential gadgets: \texttt{xzs-fusion}}
A key optimisation is fusing together consecutive dynamic applications of \mbox{\XZ{}(\S)} gadgets. Up to global phase, we have:
\begin{align*}
  \X^x\Z^z\S^s \circ \X^{x'}\Z^{z'}\S^{s'} &\simeq \X^x\Z^z (\X^{x'}\Z^{s \land x'}\S^s) \Z^{z'}\S^{s'}\\
                                           &\simeq \X^x\X^{x'}\Z^z\Z^{z'}\Z^{s \land x'}\S^s \S^{s'}\\
                                           &\simeq \X^{x \oplus x'}\Z^{z \oplus z' \oplus (s \land x')}\S^{s \oplus s'}
\end{align*}
with the first line following from the equation \(\S\X \simeq \X\Z\S\) and a case analysis. For \XZ gadgets, we have
\begin{equation*}
  \X^x\Z^z \circ \X^{x'}\Z^{z'} \simeq \X^{x \oplus x'}\Z^{z \oplus z'}
\end{equation*}
These equations are implemented through pattern rewrites.

\paragraph{Pauli propagation: \texttt{xz-commute}} As the \XZ gadget represents all (conditional) Pauli gates, it is an ideal tool for performing Pauli propagation, which commutes Pauli gates through other quantum gates, pushing them towards the end of the circuit.
The rules for commuting an \XZ gadget through a gate are specific for each gate, and are given by a ``Clifford'' trait on gate attributes, allowing this pass to apply to custom user-specified gates. This commutation data could be reused for other purposes, such as stabiliser simulation.

As an example, the case for the Hadamard gate is:
\[
  \begin{quantikz}
    &\gate{\X^x\Z^z}&\gate{\text{H}}&
  \end{quantikz}
  \rightsquigarrow
  \begin{quantikz}
    &\gate{\text{H}}&\gate{\X^z\Z^x}&
  \end{quantikz}
\]
with \X gates being converted to \Z gates when commuted past the Hadamard and \Z gates being converted to \X gates.
The case for the \CX gate is slightly more complicated; an \XZ gadget on the control wire of \CX commutes as follows:
\[
  \begin{quantikz}[row sep = small]
    &\gate{\X^x\Z^z}&\ctrl{1}&\\
    &\ghost{\X^x\Z^0}&\targ{}&
  \end{quantikz}
  \rightsquigarrow
  \begin{quantikz}[row sep = small]
    &\ctrl{1}&\gate{\X^x\Z^z}&\\
    &\targ{}&\gate{\X^x\Z^0}&
  \end{quantikz}
\]
This case highlights a difficulty of this transformation: an individual commutation rewrite for a two qubit gate can increase the number of \XZ gadgets, and a naive implementation of this compiler pass would have an exponential execution time (with respect to the length of the program).

To avoid this blow-up, we do not push each \XZ gadget all the way to the end before processing the next one. Instead, a single linear sweep commutes a gadget through at most one neighbouring gate, then greedily applies \texttt{xz-fusion} rewrites whenever adjacent \XZ gadgets are created. Immediately fusing new gadgets avoids uncontrolled duplication, ensuring linear runtime in the program length.

\paragraph{Lowering: \texttt{lower-xzs-to-select}} The last compiler pass we define for \XZ gadgets effectively reverses the action of \texttt{convert-to-xzs} and \texttt{xzs-select} by converting each \mbox{\XZ{}(\S)} gadget to a selection between classical gates.
This can be used as a first step to converting these gadgets to a more traditional representation when followed by compiler passes such as \texttt{lower-dyn-gate-to-scf}, which was introduced in the previous section.

\section{Evaluation}
\label{sec-case-studies}

We evaluate our work in two ways: application to various representative areas of quantum computing through in-depth case studies, as well as a more exhaustive benchmarking suite.
The case studies are self-contained select examples of the dynamic gates model, rather than an exhaustive list of applications.
Therefore we also evaluate our \ac{ir} on a larger benchmarking suite drawn from practice that covers the interaction of hybrid quantum-classical computing.

\subsection{Randomised Compilation}\label{sec-rand-comp}

Randomised compilation~\cite{wallmanNoiseTailoringScalable2016} is a procedure for tailoring the noise of a quantum circuit.
Quantum computation is probabilistic and it is often desirable to run a computation many times, taking the results of measurements as samples.
Instead of running the same circuit each time, randomly generated equivalent (under noiseless execution) circuits can be run, changing the noise characteristics of the computation when the results of successive runs are averaged.

The procedure acts on a circuit in the Clifford+T gate set, and begins by partitioning the gates into \emph{hard} gates (\T, \H, and \CX) and \emph{easy} gates (\(\{\X, \Y, \Z, \S, \S^\dagger\}\)).
Before each hard gate, a randomly chosen Pauli gate is applied to each of the input qubits.
After the execution of the hard gate, corrective Pauli or Phase gates are applied to the output qubits, preserving the semantics of the circuit.
We refer to the random gates added to the circuit (including the corrective operations) as \emph{padding gates}.
The padding gates are \emph{easy} gates, which can be fused with other adjacent easy gates, limiting the increase in size of the circuit due to this procedure.

The ability to simplify static Pauli and phase gates, which is possible in many existing compilers and optimisers, is not sufficient for fusing conditional padding gates.
Two ways to apply this optimisation are known:
\begin{itemize}
\item Instantiate each iteration of the circuit with its randomly generated padding gates, and then optimise the resulting programs in existing compilers individually.
\item Generate the random gates at runtime, and perform the gate merging in real time. This procedure can possibly run on the same classical control hardware used to perform error correction.
\end{itemize}
Neither option is optimal. Instantiating early duplicates work, as each iteration must go through the entire compilation pipeline. Instantiating at runtime is hardware dependent, and requires lower-level access to the control system.

The fundamental problem is that the random choice of padding gates cannot be represented as part of the quantum program, which prevents optimisation before instantiation of random variables.
The dynamic gate \ac{ir} solves this problem: the randomly generated padding gates can be encoded just like the phase flip channel in \autoref{sec-dyn-gates}, and can be fused with \XZS gadgets (see \autoref{sec-xz}) long before any random variable has its value chosen.

The padding gates are introduced by the compiler pass \texttt{randomized-comp}.
This pass matches on each hard gate, adding conditional Pauli \X and \Z gates before each input with a classical selection on a random variable and feeding the result to a dynamic gate.
It then similarly inserts dynamic gates for the correction operations to each output.
Crucially, it shares the same random variables for the padding gates before and after the operation.

Once the conditional padding gates have been added, they can be fused with the \texttt{xzs-simplify} pipeline introduced in \autoref{sec-xz}.
More precisely, this converts the padding and correction gates to \XZS gadgets, fuses the gadgets, and then lowers back to constant gates and selections.
Note that these applications need \XZS gadgets; \XZ gadgets are not expressive enough. Even though the gates generated before each hard operation are Pauli gates, placing an \X gate before a \T gate necessarily involves a Phase gate as part of the correction.
It is important not to run the \texttt{xz-commute} pass in this scenario, as it would commute the padding gates through \CX and \H gates, cancelling them and nullifying the effect of the procedure.

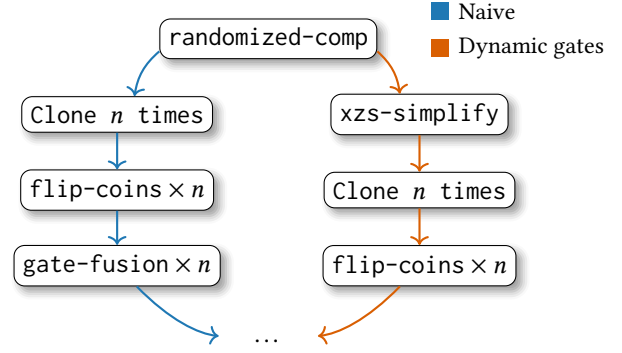
\begin{figure}
  \centering
    \begin{tikzpicture}[
    box/.style = {draw,blur shadow,shadow blur radius=0.7ex, fill=white,align=center,on grid, rounded corners},
    leftarr/.style = {color=pairedTwoDarkBlue, thick},
    rightarr/.style = {color=dark-orange, thick}
    ]
    \node[box](s1) {\texttt{randomized-comp}};
    \node[above right=0.2 and 3 of s1] {};
    \node[box, below left=1 and 2 of s1](s2) {\texttt{Clone \(n\) times}};
    \path[->, leftarr, bend right=20] (s1) edge (s2);
    \node[box, below right=1 and 2 of s1](s3) {\texttt{xzs-simplify}};
    \path[->, rightarr, bend left=20] (s1) edge (s3);
    \node[box, below=1 of s2] (s4) {\(\text{\texttt{flip-coins}} \times n\)};
    \path[->, leftarr] (s2) edge (s4);
    \node[box, below=1 of s3] (s5) {\texttt{Clone \(n\) times}};
    \path[->, rightarr] (s3) edge (s5);
    \node[box, below=1 of s4] (s6) {\(\text{\texttt{gate-fusion}} \times n\)};
    \path[->, leftarr] (s4) edge (s6);
    \node[box, below=1 of s5] (s7) {\(\text{\texttt{flip-coins}} \times n\)};
    \path[->, rightarr] (s5) edge (s7);
    \node[on grid, below=4 of s1] (end) {\dots};
    \path[->, leftarr, bend right=20, shorten >= 10pt] (s6) edge (end);
    \path[->, rightarr, bend left=20, shorten >= 10pt] (s7) edge (end);
    \matrix [below left] at (current bounding box.north east) {
      \node [fill=pairedTwoDarkBlue,label=right:\small Naive] {}; \\
      \node [fill=dark-orange,label=right:\small Dynamic gates] {}; \\
    };
  \end{tikzpicture}
  \caption{Two compiler pipelines for generating \(n\) circuits with the randomised compilation algorithm. The naive pipeline represents the state of the art, while the dynamic gate pipeline utilises dynamic gates to delay any cloning until after optimisation.}
  \label{fig-rand-comp-pipe}
\end{figure}

\begin{figure}
  \centering
  \includegraphics[width=0.8\linewidth]{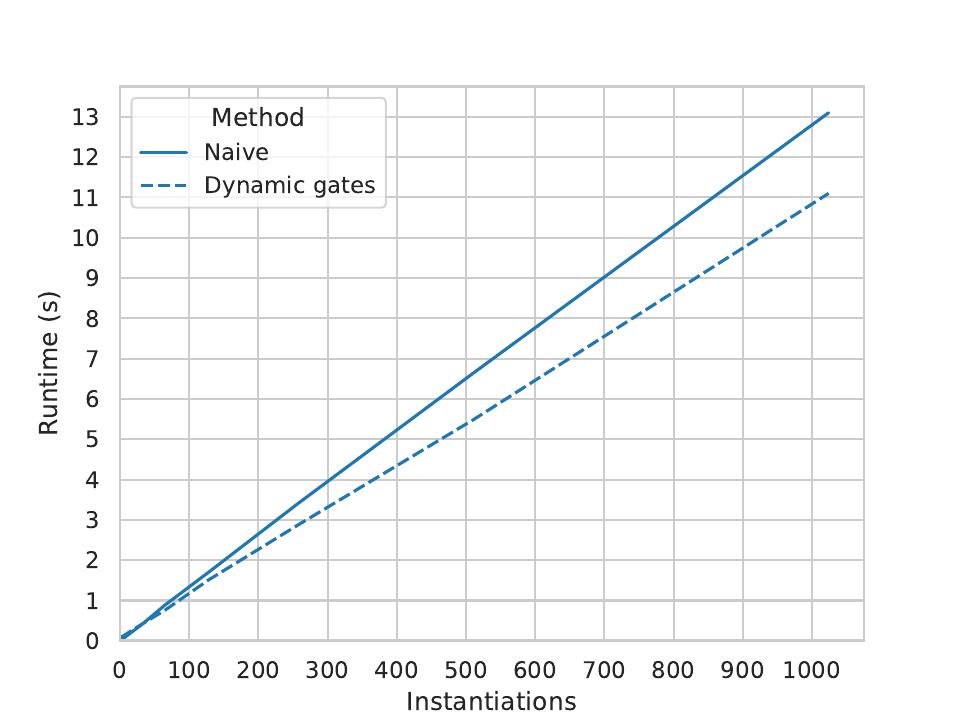}
  \caption{Time taken to generate \(n\) randomised circuits from a circuit preparing the 10-qubit GHZ state, with the naive and dynamic gate pipelines outlined in \autoref{fig-rand-comp-pipe}. With greater values of \(n\), the dynamic gate pipeline is up to 20\% faster.}
  \label{fig:rand-comp-timings}
\end{figure}

We summarise two methods of performing randomised compilation in \autoref{fig-rand-comp-pipe}:
\begin{itemize}
\item The naive pipeline represents the state of the art; \(n\) circuits are obtained by cloning the original circuit, instantiating the random variables with the \texttt{flip-coins} pass (randomly replacing each probabilistic operation by a constant), before running a pass to fuse easy gates (named \texttt{gate-fusion} in the figure) on each copy.
\item The dynamic gate pipeline represents the method detailed above, running \texttt{xzs-simplify} before cloning the circuit and instantiating random variables with \texttt{flip-coins}. In this pipeline, the fusion operation is only run once, rather than once per iteration.
\end{itemize}
We benchmark the naive pipeline against the dynamic gate pipeline on a circuit preparing a 10-qubit GHZ state, displaying the results in \autoref{fig:rand-comp-timings}. As expected, the simpler naive pipeline is faster for small numbers of iterations, but is quickly outperformed by the dynamic gate pipeline, which is up to 20\% faster for greater instantiation counts.

A secondary benefit of the dynamic gate approach is the size of the generated representation. If restricted to static quantum circuits, each individual instantiation must be stored separately, causing the representation size to scale linearly with the iteration count. Conversely, the dynamic gate representation (before instantiation) can be used to create an arbitrary number of samples. The overhead for this is small; the dynamic gate representation of applying randomised compilation to the 10-qubit GHZ state requires 237 operations, whereas the static representation requires (depending on random seeding) around 50 operations per iteration, and so is already less concise at representing 5 samples.

We believe the benefits of this approach increase the longer random variables persist in the compilation pipeline. Evaluating this is out of scope, as this work considers higher-level representations of quantum programs only. The benchmark above is effectively the worst case scenario for the dynamic gate setup, where the circuit is cloned directly after fusion -- and yet it still displays a moderate speedup.

\subsection{Quantum Error Correction}
\label{sec-qec}

Another way to combat noise in a computation is \ac{qec}; see e.g. \cite{roffeQuantumErrorCorrection2019} for a summary.
Broadly, \ac{qec} works by encoding a logical qubit as a state over multiple physical qubits, introducing redundancy.
Errors can be detected by syndrome measurements, which preserve the code-space (the states of the physical qubits corresponding to states of the logical qubit).
The results of these measurements are then passed into a decoder, a classical function that determines which error has most likely occurred.
The physical state can then be corrected with Pauli gates, returning it to the code-space and thereby correcting the errors.

In stabiliser codes, each syndrome measurement consists of \CZ and \CX gates between qubits containing the state of the logical qubit and a freshly allocated auxiliary qubit, before measuring the auxiliary qubit to get a classical bit.
The decoder could in principle be any classical computation. In the case we will discuss, it takes the form of a small boolean circuit, but for production scale \ac{qec} codes it could take the form of large matrix calculations or even neural network computations on a GPU or an FPGA.
Finally, the output of the decoder is fed into conditional Pauli gates, which we represent in the dynamic gate model as \XZ gadgets.

To preserve a quantum state, these three phases are run in a loop.
\ac{qec} protocols can only tolerate a certain number of errors while preserving the quantum state.
Noise affects idle qubits, so the protocol works better when the time between successive cycles of syndrome measurements is shorter.

One method of reducing the time between syndrome measurements is to delay the corrective operations.
The gates used in \ac{qec} are all part of the \emph{Clifford group}, the gates which normalise the Pauli gates, allowing the correction operations to be propagated past the next round of syndrome measurements.
The correction operations for successive cycles can be fused, reducing the number of quantum operations required. More importantly, they can be buffered: the next round of syndrome measurements can start before the decoder finishes execution.
This optimisation is crucial in modern \ac{qec} workflows, including recent experiments such as those performed by Google~\cite{googlequantumaiandcollaboratorsQuantumErrorCorrection2025}, that can perform up to millions of \ac{qec} cycles a second.

\begin{table}
  \centering
  \caption{The number of operations needed to present various numbers of cycles of the perfect 5-qubit \ac{qec} code in the dynamic gate \ac{ir}, both before and after application of the \texttt{xz-propagation} pipeline. The operation counts are split into quantum operations (allocations, gates, and measurements) and any other operation.}
  \label{tab:qec-size}
  \begin{tabular}{r r r r r}
    \toprule
    & \multicolumn{4}{c}{\bfseries Operations (\#)}\\
    \cmidrule(lr){2-5}
    & \multicolumn{2}{c}{\shortstack{\bfseries Before\\ \texttt{xz-propagation}}}&\multicolumn{2}{c}{\shortstack{\bfseries After\\ \texttt{xz-propagation}}}\\
    \cmidrule(lr){2-3} \cmidrule(lr){4-5}
    \textbf{Cycles (\#)} & \textbf{Quantum} & \textbf{Other} & \textbf{Quantum} & \textbf{Other}\\
    \midrule
    1&42&40&37&32\\
    10&420&400&325&500\\
    100&4200&4000&3205&5180\\
    1000&42000&40000&32005&51980\\
    \bottomrule
  \end{tabular}
\end{table}

This optimisation is performed using the \texttt{xz-propagation} pipeline.
When applied to a number of consecutive \ac{qec} cycles, the pipeline first converts each corrective gate to its \XZ gadget representation, then propagates these corrections to the end with an \texttt{xz-commute} pass, merging them together.
This method automatically modifies the results of later syndrome measurements and automatically synthesises a function to combine the results of decoding cycles.

As a prototypical case, consider the smallest error correction code capable of correcting a single qubit error: the perfect 5-qubit code~\cite{laflammePerfectQuantumError1996}.
The implementation in the dynamic gate \ac{ir} encodes all three phases of the error correction cycle, including the classical decoder.
Each phase is visually distinct within the \ac{ir}, and when the pipeline is run on two consecutive cycles, it can be easily verified by inspecting the generated \ac{ir} that only one correction phase remains.

\autoref{tab:qec-size} analyses the size of \ac{ir} required to represent this \ac{qec} code within the dynamic gate \ac{ir}. It demonstrates that the \texttt{xz-propagation} pass reduces the number of quantum operations, at the cost of a small increase in the number of classical operations.

\begin{figure}
  \centering
  \includegraphics[width=0.8\linewidth]{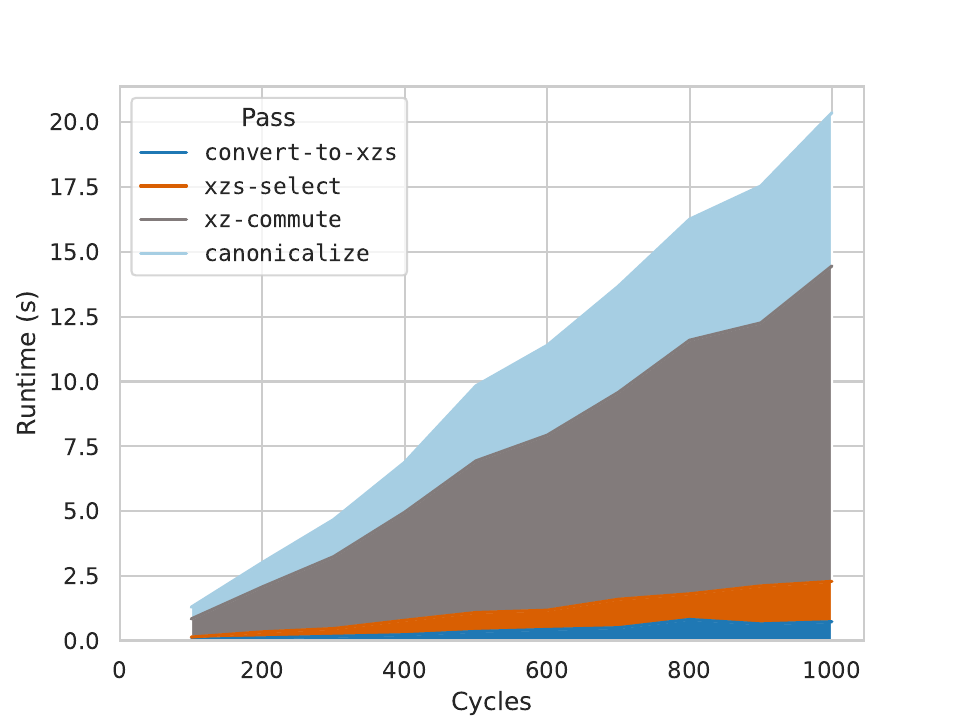}
  \caption{Duration of each pass of the \texttt{xz-propagation} pipeline when run on a varying number of cycles of the \text{5-qubit} perfect \ac{qec} code. Each pass runs in linear time with respect to the input size.}
  \label{fig:qec-timings}
\end{figure}

We further ran our pipeline on \ac{ir} representing larger numbers of consecutive \ac{qec} cycles, measuring the duration of each compiler pass in the \texttt{xz-propagation} pipeline. The results, in \autoref{fig:qec-timings}, provide evidence that the \texttt{xz-propagation} pipeline runs in linear time with respect to the number of operations in the input.

Although we only consider correction cycles for a \ac{qec} code above, and not any logical gates, the \texttt{xz-propagation} pipeline will commute corrective gates past transversal (Clifford) logical gates. We therefore conjecture that this pipeline could form part of a compiler which converts a circuit to its fault-tolerant implementation.

\subsection{Measurement-Based Quantum Computing}
\label{sec-mbqc}

An alternative to the circuit model of quantum computing is given by \ac{mbqc}~\cite{raussendorfOneWayQuantumComputer2001}.
This way of implementing unitary evolution begins by entangling the input qubits with fresh auxiliary qubits; the entire computation is then driven by performing measurements.
As quantum measurement is probabilistic, the computation is non-deterministic.
The operation can be made deterministic (unitary) again by applying corrective Pauli gates to the output qubits, conditioned on the classical outputs of the measurements.

The measurement calculus~\cite{danosMeasurementCalculus2007} gives a syntax for \ac{mbqc} programs. The normal form for \ac{mbqc} programs is given by \emph{measurement patterns}. These first allocate new qubits in the \(\ket{+} = \H\ket 0 = \frac{1}{\sqrt 2}(\ket 0 + \ket 1)\) state, entangle them (with controlled-Z gates), perform \XY-plane measurements (measurements in the basis \(\frac{1}{\sqrt 2}(\ket 0 \pm e^{i\theta}\ket 1)\) for some angle \(\theta\)), and finally correct with Pauli gates conditioned by a linear combination of the measurement results.
Consider a generic rotation, given by the unitary \(\R_\X(-\phi)\R_\Z(-\theta)\R_\X(-\lambda)\):
\[Z_{5}^{s_{1}+s_{3}}X_{5}^{s_{2}+s_{4}}[M_{4}^{\phi }]^{s_{1}+s_{3}}[M_{3}^{\theta }]^{s_{2}}[M_{2}^{\lambda }]^{s_{1}}M_{1}^{0}E_{4,5}E_{3,4}E_{2,3}E_{1,2}\]
Each operation in this pattern acts on the qubits in its subscript (with qubit 1 being the input qubit and qubit 5 being the output qubit), and \(s_n\) corresponds to the classical outcome of measuring qubit \(n\) (which is unambiguous as each qubit is measured at most once).
Each \(E_{n,m}\) is a \CZ gate on qubits \(n\) and \(m\), each \([M_n^{\psi}]^{x}\) is an \XY-plane measurement of qubit \(n\) with angle \((-1)^x\psi\), and each \(X_n^x\) or \(Z_n^x\) is a Pauli \X or \Z gate, classically conditioned on the boolean \(x\).

This pattern is in the standard CME (Correction-Measure\-ment-Entanglement) form, where  entanglement operations come before measurements, which in turn come before corrective gates.
Any well-formed pattern can be converted to this form by a standardisation procedure~\cite{broadbentParallelizingQuantumCircuits2009}.

The dynamic gate model can be used as an alternative SSA syntax for \ac{mbqc} programs.
The dynamic angle on measurement operations is accommodated by the operation \xdslinline{measurement.dyn_xy}, conditional negation of angles is explicated into an operation, and a single dynamic \XZ gadget performs the conditional \X and \Z operators. The translation of the pattern above can be found in \autoref{lst-mbqc}.

\begin{figure}
  \centering
\begin{xdsl}
\end{xdsl}
  \caption{An \ac{mbqc} program equivalent to the generic rotation \(\R_\X(-\mathtt{\textcolor[rgb]{0.00,0.20,0.40}{\%phi}})\R_\Z(-\mathtt{\textcolor[rgb]{0.00,0.20,0.40}{\%theta}})\R_\X(-\mathtt{\textcolor[rgb]{0.00,0.20,0.40}{\%lambda}})\) in our dynamic gate \ac{ir}, with input qubit \textcolor[rgb]{0.00,0.20,0.40}{\ttfamily\%q1} and output \textcolor[rgb]{0.00,0.20,0.40}{\ttfamily\%q5\_2}. Both quantum and classical dependencies are represented explicitly as SSA values.}
  \label{lst-mbqc}
\end{figure}

The \ac{ir} is not only flexible enough to represent both circuit-based programs and \ac{mbqc} programs, but can also convert from the former to the latter via local rewrites.
The transformation takes as input a quantum circuit consisting of \CZ and \(\J(\theta)\) gates, which form a universal gate set.
The generic rotation unitary \(\R_\X(-\phi)\R_\Z(-\theta)\R_\X(-\lambda)\) of our running example is given by the circuit:
\begin{center}
  \begin{quantikz}
    &\gate{\J(0)}&\gate{\J(-\lambda)}&\gate{\J(-\theta)}&\gate{\J(-\phi)}&
  \end{quantikz}
\end{center}
To allow the angles \(\lambda\), \(\theta\), and \(\phi\) to be parameters to the circuit, the operation \xdslinline{gate.dyn_j} encodes the \J gates dynamically in the \ac{ir} by taking a single angle operand.

As \ac{mbqc} programs already permit \CZ gates, the next step is to convert each \J gate to a measurement pattern:
\[
  \begin{quantikz}
    &\gate{\J(\theta)}&
  \end{quantikz}
  \rightsquigarrow
  \begin{quantikz}[wire types={q,n}]
    &&\ctrl{1}&\meter{\XY(\theta)}\\
    &\lstick{\ket{+}}&\control{}\setwiretype{q}&\gate{\X}\wire[u][1]{c}&
  \end{quantikz}
\]
The measurement is given by a dynamic \XY-plane measurement and the conditioned \X gate is given by a dynamic selection between an \X gate and an identity gate.
The pass \texttt{convert-to-cme} performs this pattern rewrite.

After this transformation, we already have a valid measurement pattern, but it is not necessarily in CME form. After all, converting composed \J gates adds a corrective Pauli gate in the middle of the program, before other \CZ gates and \XY-plane measurements.
The \texttt{xz-propagation} pass standardises the pattern by pushing the corrective Pauli gates to the end of the program (and then fusing them).
This process introduces corrective \Z gates via the commutation rules for \CZ gates, and conditionally negates angles via the commutation rules for \XY-plane measurements.

Note that this optimisation does not need extra ``signal shifting'' operations.
Previous standardisation procedures for the measurement calculus needed signal shifting to track the negations of the classical measurement results.
The explicit classical dataflow in our \ac{ir} means this is no longer required.

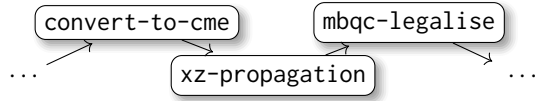
\begin{figure}[t]
  \centering
  \begin{tikzpicture}[
    box/.style = {draw,blur shadow,shadow blur radius=0.7ex, fill=white,align=center, on grid, rounded corners},
    leftarr/.style = {color=pairedTwoDarkBlue, thick},
    rightarr/.style = {color=dark-orange, thick}
    ]
    \node[on grid](s0) {\(\dots\)};
    \node[box, above right=0.7 and 1.5 of s0] (s1) {\texttt{convert-to-cme}};
    \path[->, shorten >= 3pt] (s0) edge (s1);
    \node[box, below right=0.7 and 1.8 of s1] (s2) {\texttt{xz-propagation}};
    \path[->, shorten >= 3pt] (s1) edge (s2);
    \node[box, above right=0.7 and 1.8 of s2] (s3) {\texttt{mbqc-legalise}};
    \path[->, shorten >= 3pt] (s2) edge (s3);
    \node[on grid, below right=0.7 and 1.5 of s3](s4) {\(\dots\)};
    \path[->, shorten >= 3pt] (s3) edge (s4);
  \end{tikzpicture}
  \caption{A compiler pipeline to convert \CZ/\J circuits to \ac{mbqc} programs within the dynamic gate \ac{ir}.}
  \label{fig-mbqc}
\end{figure}

Finally, the \texttt{mbqc-legalise} pass reorders the operations to result in a valid CME pattern, leaving the dataflow graph of the program unchanged. The full pipeline is summarised in \autoref{fig-mbqc}.

\subsection{Benchmarking Suite}\label{sec-benchmarking}

To gather more general metrics of our \ac{ir}, we curated a new benchmarking suite of hybrid quantum-classical algorithms. To ensure that the programs we have picked are representative, we followed the overview of hybrid classical-quantum algorithms given by \citeauthor{ellaQuantumclassicalProcessingBenchmarking2023}~\cite[Fig. 1]{ellaQuantumclassicalProcessingBenchmarking2023}. Although this previous analysis focusses on pulse-level programs, we surveyed the papers they cited, and after filtering for gate-level algorithms arrived at the programs in \autoref{tab:benchmark}.

\begin{table}[t]
  \setlength{\tabcolsep}{0em}
  \noindent\begin{tabular}{p{\linewidth}}
    \toprule
    \textbf{Teleport}: Constant-depth teleportation protocol~\cite[Fig. 1]{devulapalliQuantumRoutingTeleportation2024}.\\
    \textbf{Prep}: State preparation of logical plus state~\cite{foss-feigExperimentalDemonstrationAdvantage2023}.\\
    \textbf{RUS}: Three Repeat-Until-Success circuits for \(\writegate{V}_3\)~\cite[Fig. 1]{adamRepeatUntilSuccessNondeterministicDecomposition2014}.\\
    \textbf{QML}: Conditional gearbox circuit from a quantum neural network~\cite[Fig. 1]{moreiraRealizationQuantumNeural2023}.\\
    \textbf{IPE}: Iterative phase estimation~\cite[Fig. 1 (Bottom)]{corcolesExploitingDynamicQuantum2021}.\\
    \textbf{RWPE}: Random walk phase estimation~\cite[Algorithm 1]{granadeUsingRandomWalks2022}.\\
    \textbf{MBQC-Rot}: General \ac{mbqc} rotation gate~\cite{raussendorfOneWayQuantumComputer2001}.\\
    \textbf{MBQC-CX}: \ac{mbqc} \CX gate~\cite{raussendorfOneWayQuantumComputer2001}.\\
    \textbf{QEC-Adap}: Adaptive \ac{qec} cycle~\cite[Fig. 17]{ryan-andersonRealizationRealTimeFaultTolerant2021}.\\
    \textbf{QAOA}: Max-cut Quantum Approximate Optimisation Algorithm. See e.g.~\cite{farhiQuantumApproximateOptimization2014,cerezoVariationalQuantumAlgorithms2021}.\\
    \bottomrule
  \end{tabular}
  \caption{Hybrid Benchmark Suite. Implementations in our \ac{ir} are contained in the artifact~\cite{Artifact}.}
  \label{tab:benchmark}
\end{table}

\begin{figure*}[t]
  \centering
  \begin{subcaptionblock}{0.33\linewidth}
    \includegraphics[width=\linewidth]{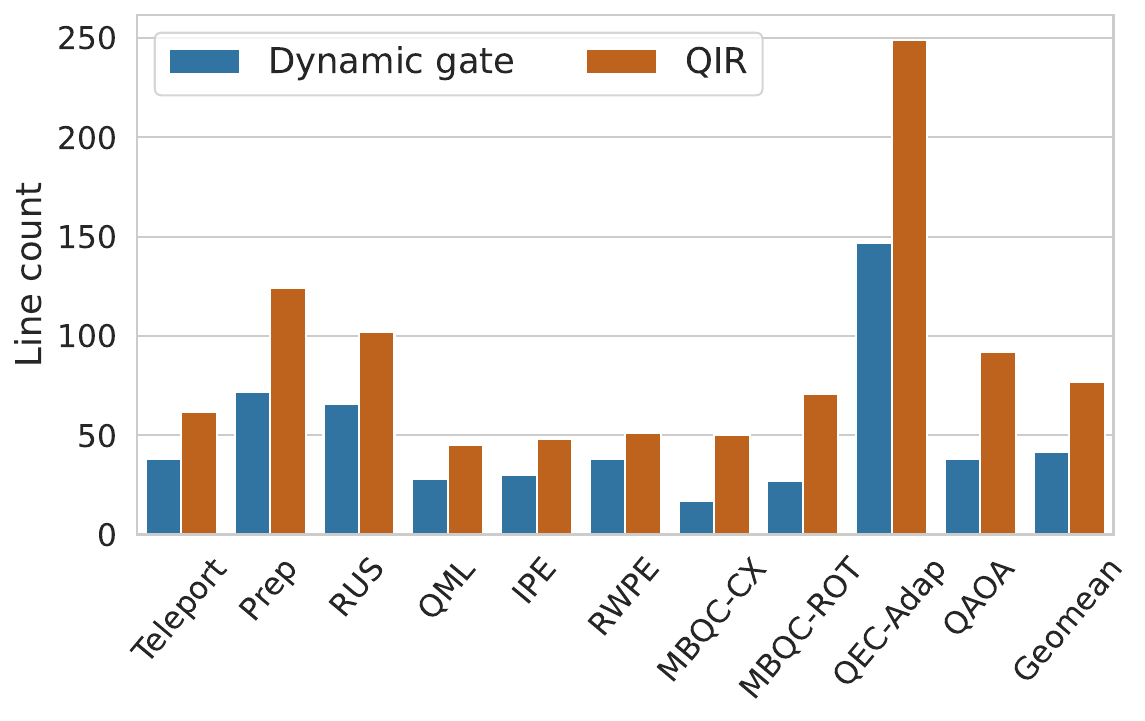}
    \caption{Line count}
  \end{subcaptionblock}
  \begin{subcaptionblock}{0.33\linewidth}
    \includegraphics[width=\linewidth]{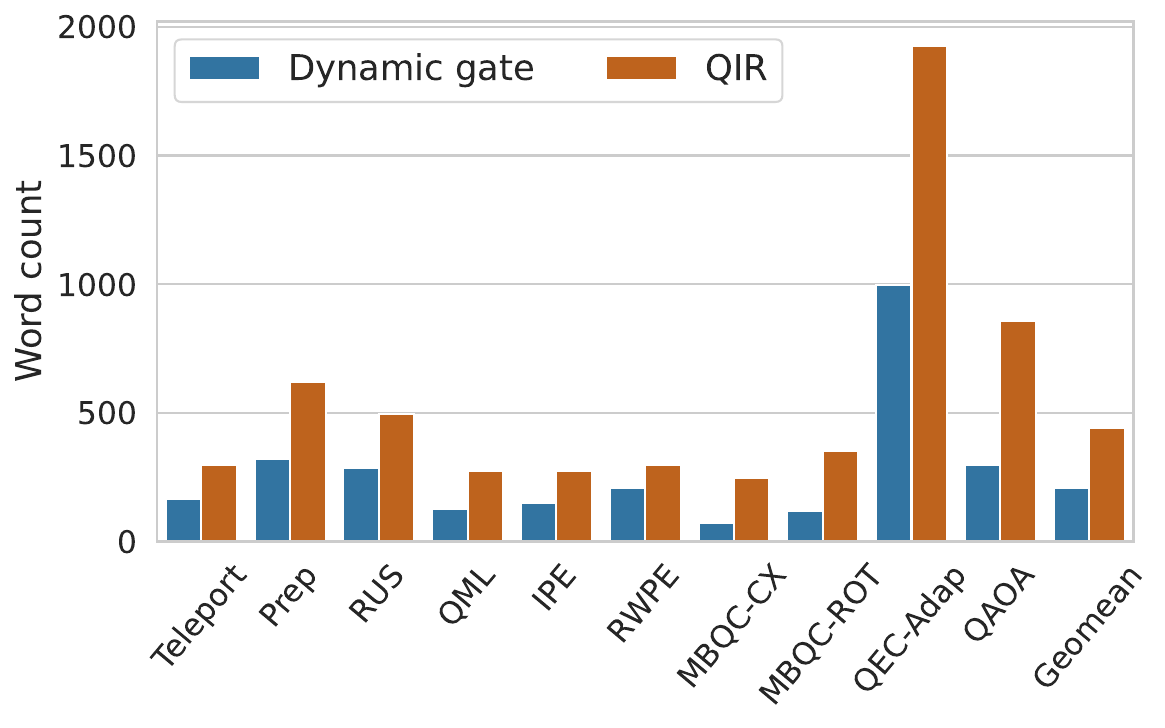}
    \caption{Word count}
  \end{subcaptionblock}
  \begin{subcaptionblock}{0.33\linewidth}
    \includegraphics[width=\linewidth]{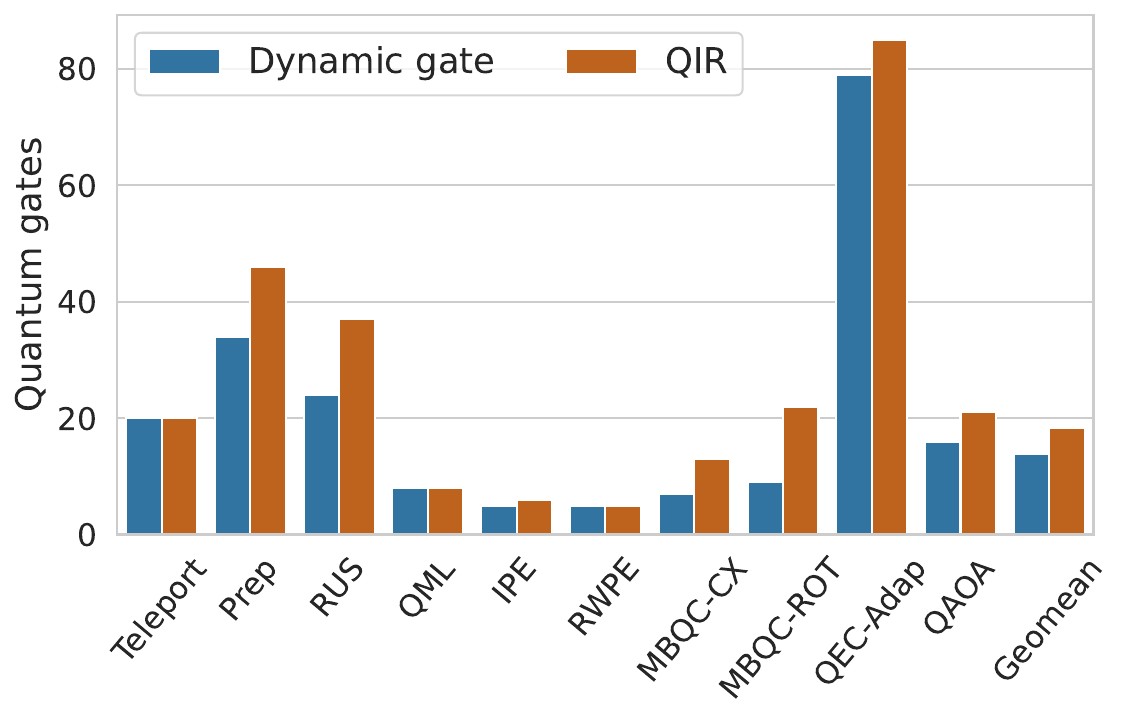}
    \caption{Quantum operation count}
  \end{subcaptionblock}
  \begin{subcaptionblock}{0.33\linewidth}
    \includegraphics[width=\linewidth]{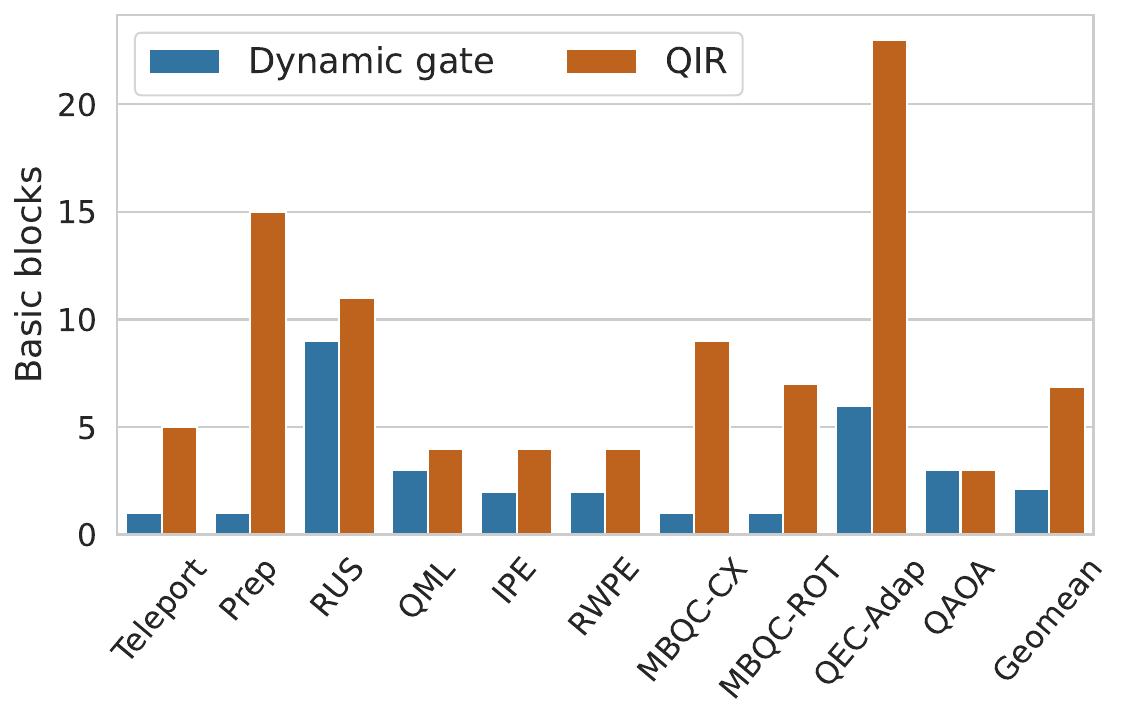}
    \caption{Basic block count}
  \end{subcaptionblock}
  \begin{subcaptionblock}{0.33\linewidth}
    \includegraphics[width=\linewidth]{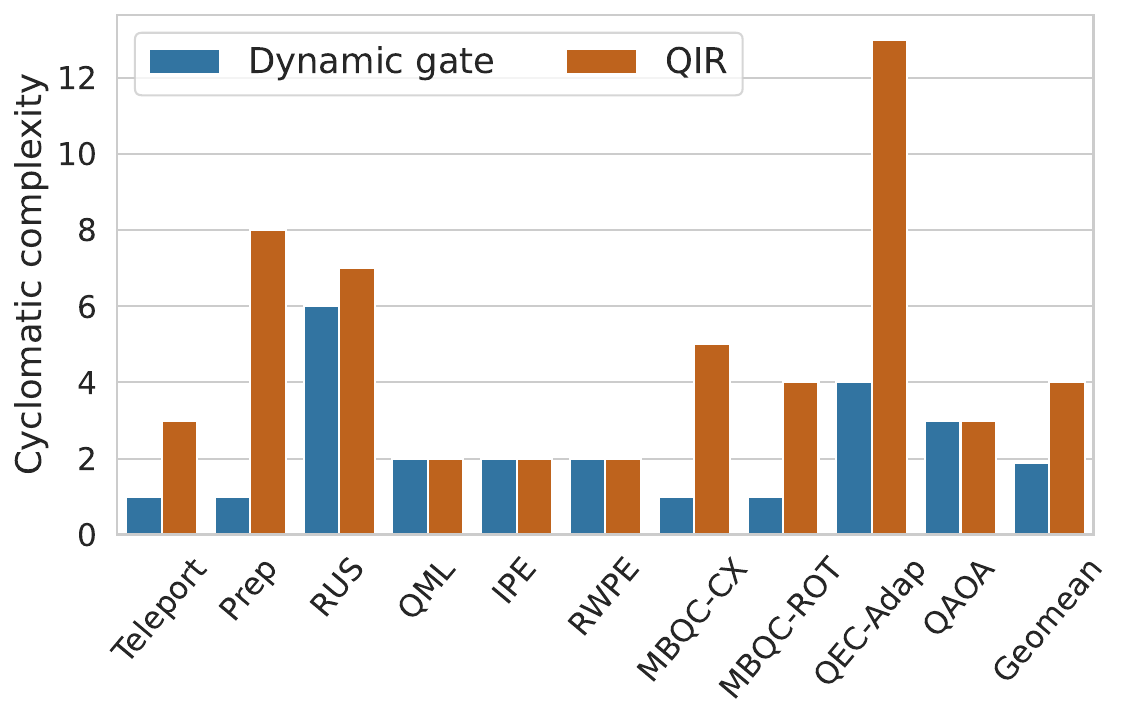}
    \caption{Cyclomatic complexity}
  \end{subcaptionblock}
  \begin{subcaptionblock}{0.33\linewidth}
    \includegraphics[width=\linewidth]{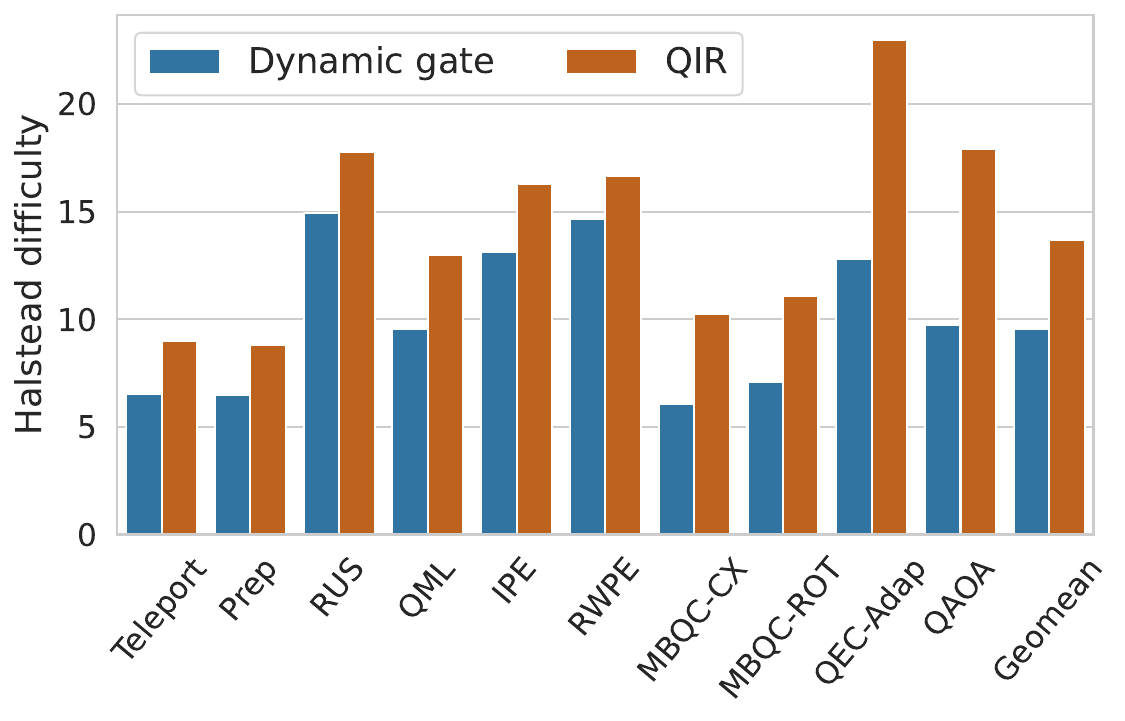}
    \caption{Halstead difficulty}
  \end{subcaptionblock}
  \caption{Numerical analysis on the benchmark suite. The Dynamic Gate \ac{ir} is more concise than QIR, and is able to rely less on (unstructured) control flow operations, reducing the overall complexity of programs.}
  \label{fig:benchmark}
\end{figure*}

While not a comprehensive list of all algorithms, the benchmark contains many of the constructions typically found in hybrid programs, including angle-parameterised gates, conditional gate application, and classical loops. It includes programs from different quantum computing paradigms, such as measurement driven computation, NISQ variational algorithms, and building blocks of fault-tolerant computation. We chose more focussed, concise programs, with line counts varying from 17 to 147, easing the manual implementation and tuning of these programs for other systems in the future.

This benchmarking suite is used to compare our \ac{ir} to QIR. To obtain QIR programs, we systematically lowered each program in the benchmarking suite. First, \texttt{qssa} operations were converted to their \texttt{qref} counterparts. Then, dynamic gates were lowered to structured control flow operations, which were further lowered to branching operations. Our gate operations then had to be converted to QIR style gate invocations. Finally, we used MLIR passes to reduce the classical components of the program to LLVM-IR. We lowered to the quantum instruction set supported by qir-runner~\cite{QIRRunner}, an official QIR alliance project.

We performed six analyses of the programs, the results of which are graphed in \autoref{fig:benchmark}. The analyses chosen require no human input and are purely numerical metrics.

The line and word counts of each program provide an initial measure of how verbose each program is, with the line count correlating strongly to the total number of operations in each program. We were further able to count the number of quantum operations in each program, which we define to be \texttt{qssa} operations for our \ac{ir}, and calls to functions of the form \verb|@__quantum__qis*| for QIR. The difference between the representations in this metric arises from the representation of conditional gates, as well as our \ac{ir} allowing non-zero allocation and measurement in bases other than the Z basis.

We further included counts of the number of basic blocks in each program as well as cyclomatic complexity~\cite{McCabeComplexity}, to measure the complexity of the control flow of each program. Lastly, we include the following metric \(H\) of the complexity of a program due to \citeauthor{HalsteadElements}~\cite{HalsteadElements}:
\[ H = \frac{\text{\# of unique operations}}{2} \times \frac{\text{Total \# of operands}}{\text{\# of unique operands}}\]
Here, we define operands as in MLIR/LLVM, and consider two operations to be equal if they are identical other than their operands (in particular letting applications of different gates be distinct operations).

Our \ac{ir} outperformed QIR in every analysis. When comparing the geometric means over each benchmark (listed as the final entry in each plot of \autoref{fig:benchmark}), dynamic gates used 47\% fewer lines, 52\% fewer words, 24\% fewer quantum operations, 69\% fewer basic blocks, had 52\% less cyclomatic complexity, and 30\% less Halstead difficulty.

\section{Related and Future Work}

In this section, we analyse some of the differences between the \ac{ir} presented in this paper with other previously introduced quantum \acp{ir}. We then review other work that could form the basis for different gadgets, or that we otherwise believe could be represented well by the dynamic gate framework. We follow this discussion with some further suggestions of future work and potential additions to the \ac{ir}.

\subsection{Related Quantum Intermediate Representations}

There exist two dominant textual representations for quantum programs: OpenQASM and QIR. OpenQASM~\cite{crossOpenQASM3Broader2022} closely corresponds to the representation of quantum programs in the Qiskit~\cite{javadi-abhariQuantumComputingQiskit2024} quantum toolkit. Version 2 of OpenQASM was restricted to quantum circuits with a fixed number of qubits, though version 3 has added a limited ability to represent classical computations. Conversely, QIR~\cite{QIRSpec2021} is built on top of LLVM~\cite{lattnerLLVMCompilationFramework2004}, and inherits all its classical functionality, but quantum gates are applied via opaque function calls with little interaction between the classical and quantum components of the program. In both these formats, qubits have reference semantics, unlike the value semantics common in modern compilers.

Other quantum \acp{ir} have been proposed within the MLIR framework. \citeauthor{mccaskeynguyen:qir}~\cite{mccaskeynguyen:qir} introduce an MLIR dialect for representing quantum programs, leveraging the tooling present in MLIR to define a translation down to QIR, while remaining expressive enough to provide lowerings from other toolkits to this dialect. Like QIR, qubits in this dialect are given by reference semantics. The first MLIR quantum dialect using value semantics was QSSA~\cite{peduriQSSASSAbasedIR2022}. QIRO~\cite{ittahQIROStaticSingle2022}, which is the basis of the IR used in the Catalyst compiler~\cite{ittahCatalystPythonJIT2024} (distributed as part of Pennylane~\cite{bergholmPennyLaneAutomaticDifferentiation2022}), introduces parallel quantum dialects; one with reference semantics for qubits and one with value semantics, leveraging that linearity analysis is easier within the non-SSA dialect, and optimisations are easier within the SSA dialect.

Ensemble IR~\cite{wawdhaneEnsembleIRConciseRepresentation2025} provides MLIR dialects for representing collections (or ensembles) of related quantum circuits, as in the randomised compilation example of \autoref{sec-rand-comp}. Similar to the current work, it allows gates to be specified via an extensible set of strings, rather than having fixed operations for each gate, and attempts to perform optimisations and transformations at the hybrid level rather than circuit level. It is more specialised to probabilistic scenarios than a general hybrid setting, containing primitive operations for certain probabilistic tasks such as sampling from distributions. Future work could study how these dialects could be used in conjunction with our own, given their similar approaches and setting.

Quantinuum recently introduced t\(|\)ket\(\rangle\) 2, extending t\(|\)ket\(\rangle\) 1~\cite{Sivarajah_TKET_A_Retargetable_2020} to hybrid quantum-classical programs. It is built on HUGR~\cite{koch2025hugrquantumclassicalintermediaterepresentation}, a custom-built library for creating quantum-classical representations. A possible avenue for future work is looking at the feasibility of introducing the framework, operations, and accompanying translations to this library.

\subsection{Quantum Gadgets for Optimisation}

A range of quantum gadgets already exist in the literature for optimising quantum circuits. Phase polynomials~\cite{amyPolynomialTimeTDepthOptimization2014} are a gadget for representing combinations of \CX and \T gates, primarily used to reduce the \T-count of a circuit.

The ZX calculus~\cite{coeckeInteractingQuantumObservables2011} is a graph-based method of representing linear maps. It introduces quantum gadgets called red and green spiders, and it has been shown these gadgets can be used to optimise quantum circuits~\cite{kissingerReducingNumberNonClifford2020}. As future work it would be interesting to explore adding ZX spiders to the dynamic gate model, which is nontrivial because the ZX calculus makes no distinction between the inputs and outputs of spiders.

The t\(|\)ket\(\rangle\) 1~\cite{Sivarajah_TKET_A_Retargetable_2020} quantum circuit compiler makes use of two quantum gadgets inspired by the ZX calculus: phase gadgets, and their generalisation Pauli gadgets. These gadgets are used like the \XZ gadgets in this paper: subcircuits can be represented by phase gadgets, which can be optimised before translating back to a traditional circuit representation.

Future work could investigate which of these are suitable for implementation as dynamic gates, allowing them to be used in the optimisation of hybrid programs rather than static quantum circuits.

\subsection{Other Future Directions}

Much of the current work focusses on optimisations concerning Clifford gates, with the \XZ gadgets representing Pauli gates and the \XZS gadgets adding the Clifford phase gate. The optimisations in this paper have utility outside of Clifford programs, such as the randomised compilation in \autoref{sec-rand-comp} on Clifford+\T circuits and the universal \ac{mbqc} programs of \autoref{sec-mbqc}. In the future it would be good to explore optimisations which target hybrid non-Clifford operations. Adapting phase polynomials (see the previous section) to the dynamic gate framework could be a way to achieve this.

The current work does not discuss subprocedures or subcircuits. While these can be represented with MLIR's \texttt{func} dialect, a more complete solution would be able to interact with conversions between reference and value-based semantics, Pauli propagation, and inverting unitaries. An interesting direction for further research is to explore implementing a \texttt{subcircuit} operation producing a gate value from a circuit, which could then be applied dynamically.

A different direction to extend the current work is to add quantum data types other than qubits. The addition of quantum registers has been explored in other quantum \acp{ir}, with QIRO and QSSA both introducing custom quantum register types with operations for manipulating them. Note that the MLIR \texttt{tensor} type is immutable, making it difficult to use on qubits with value-semantics. Other work~\cite{vaxQmodExpressiveHighLevel2025,yuanTowerDataStructures2022} has explored adding higher-level data types to quantum languages, which will be necessary to represent in future quantum \acp{ir}.

Finally, while quantum noise channels were used to motivate dynamic gates, their use within a noise-aware compilation pipeline remains relatively unexplored. Natively representing noise channels with dynamic gates should in particular work for the classical simulation of noisy circuits, where it should be possible to optimise the noisy circuit before instantiating the noise for a particular sample, much like in our randomised compilation case study.

\begin{acks}                            
  The authors would like to thank Chris Vasiladiotis for feedback on earlier drafts of this paper.
  This work was funded by EPSRC Grant \grantnum{Roarq}{EP/W032635/1}, \grantsponsor{Roarq}{Robust and Reliable Quantum Computing}.
\end{acks}

\bibliography{references}

\appendix
\clearpage
\onecolumn
\section{Dynamic Gate IR Specification}
\label{tab:dynamic-gate-ir}
Here we list the attributes, types, and operations of the Dynamic Gate IR, which are to be combined with common MLIR dialects (e.g. arith and builtin). Crucially, the IR is open in nature; a user can define new gate attributes or operations for creating gates, and combine them with the operations listed below.

\begin{center}
  \begin{tabular}{llll}
    \toprule
    \textbf{Types} &&&\\
    \midrule
    \xdslinline{!qu.bit}&\multicolumn{3}{l}{Qubit type}\\
    \xdslinline{!gate.type<£\(n\)£>}&\multicolumn{3}{l}{Type of \(n\)-qubit gates}\\
    \xdslinline{!measurement.type<£\(n\)£>}&\multicolumn{3}{l}{Type of \(n\)-qubit measurements}\\
    \xdslinline{!angle.type}&\multicolumn{3}{l}{Type of angles}\\
    \midrule
    \textbf{Attributes}&&&\\
    \midrule
    Angle attribute&\(\theta\)&:=&\xdslinline{#angle.attr<£\(x\)£>} \hspace{0pt plus 1filll} for \(0 \leq x < 2\pi\)\\
    Gate attributes&\(g\)&:=&\xdslinline{#gate.id<£\(n\)£>}\(\ \mid\ \)\xdslinline{#gate.x}\(\ \mid\ \)\xdslinline{#gate.y}\(\ \mid\ \)\xdslinline{#gate.z}\(\ \mid\ \)\xdslinline{#gate.h}\(\ \mid\)\\
                   &&&\xdslinline{#gate.s}\(\ \mid\ \)\xdslinline{#gate.s_dagger}\(\ \mid\ \)\xdslinline{#gate.t}\(\ \mid\ \)\xdslinline{#gate.t_dagger}\(\ \mid\)\\
                   &&&\xdslinline{#gate.rx<£\(\theta\)£>}\(\ \mid\ \)\xdslinline{#gate.ry<£\(\theta\)£>}\(\ \mid\ \)\xdslinline{#gate.rz<£\(\theta\)£>}\(\ \mid\ \)\xdslinline{#gate.j<£\(\theta\)£>}\(\ \mid\)\\
                   &&&\xdslinline{#gate.rzz<£\(\theta\)£>}\(\ \mid\ \)\xdslinline{#gate.cx}\(\ \mid\ \)\xdslinline{#gate.cz}\(\ \mid\ \)\xdslinline{#gate.toffoli}\\
    Measurement attributes&\(m\)&:=&\xdslinline{#measurement.comp_basis}\(\ \mid\ \)\xdslinline{#measurement.x_basis}\(\ \mid\)\\
                   &&&\xdslinline{#measurement.xy<£\(\theta\)£>}\\
    Allocation attributes&\(a\)&:=&\xdslinline{#qu.zero}\(\ \mid\ \)\xdslinline{#qu.plus}\\
    \midrule
    \textbf{Operations} &\multicolumn{3}{l}{\textbf{Type}}\\
    \midrule
    \xdslinline{qu.alloc<£\(a\)£>}&\multicolumn{3}{l}{() -> \xdslinline{!qu.bit}\(^n\)\hspace{0pt plus 1filll}where \(\mathtt{qubits}(a) = n\)}\\
    \xdslinline{qssa.gate<£\(g\)£>}&\multicolumn{3}{l}{\xdslinline{!qu.bit}\(^n\) -> \xdslinline{!qu.bit}\(^n\)\hspace{0pt plus 1filll}where \(\mathtt{qubits}(g) = n\)}\\
    \xdslinline{qssa.dyn_gate}&\multicolumn{3}{l}{(\xdslinline{!gate.type<£\(n\)£>}, \xdslinline{!qu.bit}\(^n\)) -> \xdslinline{!qu.bit}\(^n\)}\\
    \xdslinline{qssa.measure<£\(m\)£>}&\multicolumn{3}{l}{\xdslinline{!qu.bit}\(^n\) -> \xdslinline{i1}\(^n\)\hspace{0pt plus 1filll}where \(\mathtt{qubits}(m) = n\)}\\
    \xdslinline{qssa.dyn_measure}&\multicolumn{3}{l}{(\xdslinline{!measurement.type<£\(n\)£>}, \xdslinline{!qu.bit}\(^n\)) -> \xdslinline{i1}\(^n\)}\\
    \xdslinline{qref.gate<£\(g\)£>}&\multicolumn{3}{l}{\xdslinline{!qu.bit}\(^n\) -> ()\hspace{0pt plus 1filll}where \(\mathtt{qubits}(g) = n\)}\\
    \xdslinline{qref.dyn_gate}&\multicolumn{3}{l}{(\xdslinline{!gate.type<£\(n\)£>}, \xdslinline{!qu.bit}\(^n\)) -> ()}\\
    \xdslinline{qref.measure<£\(m\)£>}&\multicolumn{3}{l}{\xdslinline{!qu.bit}\(^n\) -> \xdslinline{i1}\(^n\)\hspace{0pt plus 1filll}where \(\mathtt{qubits}(m) = n\)}\\
    \xdslinline{qref.dyn_measure}&\multicolumn{3}{l}{(\xdslinline{!measurement.type<£\(n\)£>}, \xdslinline{!qu.bit}\(^n\)) -> \xdslinline{i1}\(^n\)}\\
    \xdslinline{gate.constant £\(g\)£}&\multicolumn{3}{l}{() -> \xdslinline{!gate.type<£\(n\)£>}\hspace{0pt plus 1filll}where \(\mathtt{qubits}(g) = n\)}\\
    \xdslinline{gate.xz}&\multicolumn{3}{l}{(\xdslinline{i1}, \xdslinline{i1}) -> \xdslinline{!gate.type<1>}}\\ \xdslinline{gate.xzs}&\multicolumn{3}{l}{(\xdslinline{i1}, \xdslinline{i1}, \xdslinline{i1}) -> \xdslinline{!gate.type<1>}}\\
    \xdslinline{gate.dyn_rx}&\multicolumn{3}{l}{\xdslinline{!angle.type} -> \xdslinline{!gate.type<1>}}\\
    \xdslinline{gate.dyn_ry}&\multicolumn{3}{l}{\xdslinline{!angle.type} -> \xdslinline{!gate.type<1>}}\\
    \xdslinline{gate.dyn_rz}&\multicolumn{3}{l}{\xdslinline{!angle.type} -> \xdslinline{!gate.type<1>}}\\
    \xdslinline{gate.dyn_j}&\multicolumn{3}{l}{\xdslinline{!angle.type} -> \xdslinline{!gate.type<1>}}\\
    \xdslinline{gate.dyn_rzz}&\multicolumn{3}{l}{\xdslinline{!angle.type} -> \xdslinline{!gate.type<2>}}\\
    \xdslinline{measurement.constant £\(m\)£}&\multicolumn{3}{l}{() -> \xdslinline{!measurement.type<£\(n\)£>}\hspace{0pt plus 1filll}where \(\mathtt{qubits}(m) = n\)}\\
    \xdslinline{measurement.dyn_xy}&\multicolumn{3}{l}{\xdslinline{!angle.type} -> \xdslinline{!measurement.type<1>}}\\
    \xdslinline{angle.constant £\(\theta\)£}&\multicolumn{3}{l}{() -> \xdslinline{!angle.type}}\\
    \xdslinline{angle.add}&\multicolumn{3}{l}{(\xdslinline{!angle.type}, \xdslinline{!angle.type}) -> \xdslinline{!angle.type}}\\
    \xdslinline{angle.scale}&\multicolumn{3}{l}{(\xdslinline{!angle.type}, \xdslinline{f64}) -> \xdslinline{!angle.type}}\\
    \xdslinline{angle.negate}&\multicolumn{3}{l}{\xdslinline{!angle.type} -> \xdslinline{!angle.type}}\\
    \xdslinline{angle.cond_negate}&\multicolumn{3}{l}{(\xdslinline{i1}, \xdslinline{!angle.type}) -> \xdslinline{!angle.type}}\\
    \xdslinline{prob.bernoulli £\(f\)£}&\multicolumn{3}{l}{() -> \xdslinline{i1}\hspace{0pt plus 1filll}where \(0 \leq f \leq 1\)}\\
    \xdslinline{prob.uniform £\(T\)£}&\multicolumn{3}{l}{() -> \(T\)\hspace{0pt plus 1filll}where \(T\) is an integer type}\\
    \bottomrule
  \end{tabular}
\end{center}

\end{document}